\documentclass[%
 reprint,
 amsmath,amssymb,
 aps,
]{revtex4-2}

\usepackage{graphicx}
\graphicspath{{./figures/}}
\usepackage{dcolumn}
\usepackage{bm}
\usepackage{url}

\begin{document}

\preprint{APS/123-QED}

\title{The Demise of Single-Authored Publications in Computer Science:\\ A Citation Network Analysis}

\author{Brian K. Ryu}
 \altaffiliation[\url{bryu@stanford.edu}]{}
\affiliation{%
 Department of Chemical Engineering\\
 Stanford University, Stanford, CA, 94305
}%




\date{\today}

\begin{abstract}
In this study, I analyze the DBLP bibliographic database to study role of single author publications in the computer science literature between 1940 and 2019. I examine the demographics and reception by computing the population fraction, citation statistics, and PageRank scores of single author publications over the years. Both the population fraction and reception have been continuously declining since the 1940s. The overall decaying trend of single author publications is qualitatively consistent with those observed in other scientific disciplines, though the diminution is taking place several decades later than those in the natural sciences. Additionally, I analyze the scope and volume of single author publications, using page length and reference count as first-order approximations of the scope of publications. Although both metrics on average show positive correlations with citation count, single author papers show no significant difference in page or reference counts compared to the rest of the publications, suggesting that there exist other factors that impact the citations of single author publications.

\end{abstract}

\maketitle

\section{Introduction}
Trends in executing and publishing scientific research have been constantly changing throughout history \cite{egghe1990introduction}. For example, the body of scientific literature is slowly but consistently becoming more interconnected and interdisciplinary. The study by Porter and Rafols which analyzed various ares of scientific literature has found that the importance of monodisciplinary publications in all academic fields have decreased over the past four decades \cite{porter2009science}. While certain fields such as math still remain more monodisciplinary than other interdisciplinary subject areas such as materials science, the authors showed that the increase in subject interconnectedness and mean number of authors has been a universal transition exhibited by all disciplines. A detailed study of citation statistics of Physical Review journals up to 2004 by Redner noted the citation patterns of individual publications \cite{redner2004citation}. Redner showed that a paper typically receives the most citations during the first several years after publication, followed by a rapid decay. However, that there exist occasional \textit{revivals of old classics} or \textit{sleeping beauties} \cite{van2004sleeping} that defy the typical monotonic decline of citation probability over time, as well as \textit{major discoveries} that cause a sharp spike in a recognized work. Overall, the evolution of citation networks involve dynamic features and trends at both the holistic and individual levels.

A key trend in the history of publications that persists today is the decline of single author publications. The wane of single author publications and wax of co-authored publications owes to the fundamental shift towards research networks and collaboration in the geography of science, as pointed out by Adams \cite{adams2012collaborations}. In 1963, Price noted that most early papers from \textit{Chemical Abstracts} were authored by sole researchers but the emergence of co-authored papers in the 1950s and 1960s have led to a sharp decline in the proportion of single author publications \cite{de1965little}. Price further predicted that if the decline persists, single author papers would become extinct by 1980. Although single author papers are still being published in present day, similar decaying trends have been observed across all scientific disciplines. Abt pointed out that over 90\% of all astronomical papers from the early 20th century were single authored, but most publications in astronomy from 2005 were co-authored with 4.9 authors per paper on average \cite{abt1981some, abt2007future}. Similar trends were observed from economics. Hudson showed that multi-authorship in leading economic journals rose from less than 10\% in 1950 to over 50\% in 1990 \cite{hudson1996trends}. A recent study by Kuld and O'Hagan found that co-authored publications comprised over 70\% of works published in the top 255 economics journals in 2014 \cite{kuld2018rise}. Scientific disciplines that frequently involve collaborative efforts such as ecology have experienced more dramatic declines of single author publications \cite{barlow2018extinction}. The study by Barlow \textit{et al.} has shown that as of 2017, the proportion of single author works published in the \textit{Journal of Applied Ecology} is less than 5\%. Moreover the average acceptance rate for manuscripts with one author is less than half of that with two or more authors. Publishing alone in the 21st century has become much difficult than in the 20th century.

The present study analyzes the decline of the proportion and citation impact of single authored publications in the computer science literature. Computer science is a young field in the history of publications. Since its inception in the 1940s, an explosive rise in the number of publications in computer science has led to interest in bibliographic studies of the computer science literature \cite{fernandes2014authorship, solomon2009programmers, heilig2014scientometric}. These studies have observed an emergence of collaborative studies, similar to those in other scientific disciplines \cite{franceschet2011collaboration}. 
Yet, the role and decline of single authored publications in the computer science literature has not received much attention. To obtain a comprehensive understanding of the current state of the literature, I analyze over 4 million publications and their citation relations from the entire DBLP bibliographic database. I first present and discuss general co-citation trends and compare them to those from other scientific disciplines. Then, I discuss the role and impact of single author publications via citation relations. The analysis leads to the conclusion that single authored publications in compute science computer science are still important today yet gradually and consistently losing importance.

\section{Methods}
The full DBLP citation network database, acquired by Arnetminer and available at \url{https://aminer.org/citation} was used to study the network of computer science literature \cite{tang2008arnetminer}. The DBLP bibliography includes all major journal articles and arXiv computer science pre-prints, as well as conference proceedings. In computer science, both journal publications and conference proceedings are prestigious and serve as indispensable means of disseminating scientific research \cite{meyer2009research}. The use of the DBLP database has been common in other bibliometric studies of computer science literature \cite{cavero2014computer, elmacioglu2005six, fernandes2017evolution}. The citation network contains 4,107,340 scientific publications (nodes) and 36,624,464 citation relationships (edges). The dataset was extracted on May 5th, 2019.

In addition to basic citation relations, the dataset contains detailed information pertaining to each publication such as author names and IDs, publication year, number of pages, and number of citations, which are analyzed in this work. The number of citations includes counts from publications outside of the computer science field that are not accounted for in the DBLP database. All but two publications in the dataset included these attributes, and the two publications without these attributes were excluded during analysis. Therefore, the studied dataset accounts for over 99.999\% of the entire DBLP database. 


All average quantities reported here are calculated as arithmetic means. While some studies report the adequacy of geometric means when evaluating skewed metrics such as citations \cite{thelwall2017three}, general trends observed from a large dataset comprising over 3 million samples should not be affected. In reporting time-evolution of measurements, I do not report measurements for publications preceding 1940 due to the small number of publication counts. Metrics for publications from 2019 are not reported as well due to incomplete statistics for publications from 2019. However, citations from these publications are included in computing citations of preceding works.

\section{Results}
The analysis of the entire DBLP database enables the study of time-evolution of trends. In this section, I first present results on the general trends of publication statistics and authorships, followed by a more detailed investigation on the impact of single-authored papers. Finally, I compare the volume and scope of single author publications to those of all publications in the dataset to discuss the quality of single authored publications.

\subsection{General Citations and Authorship Facts}
\begin{figure}[h]
\centering
\includegraphics[width=8.5cm]{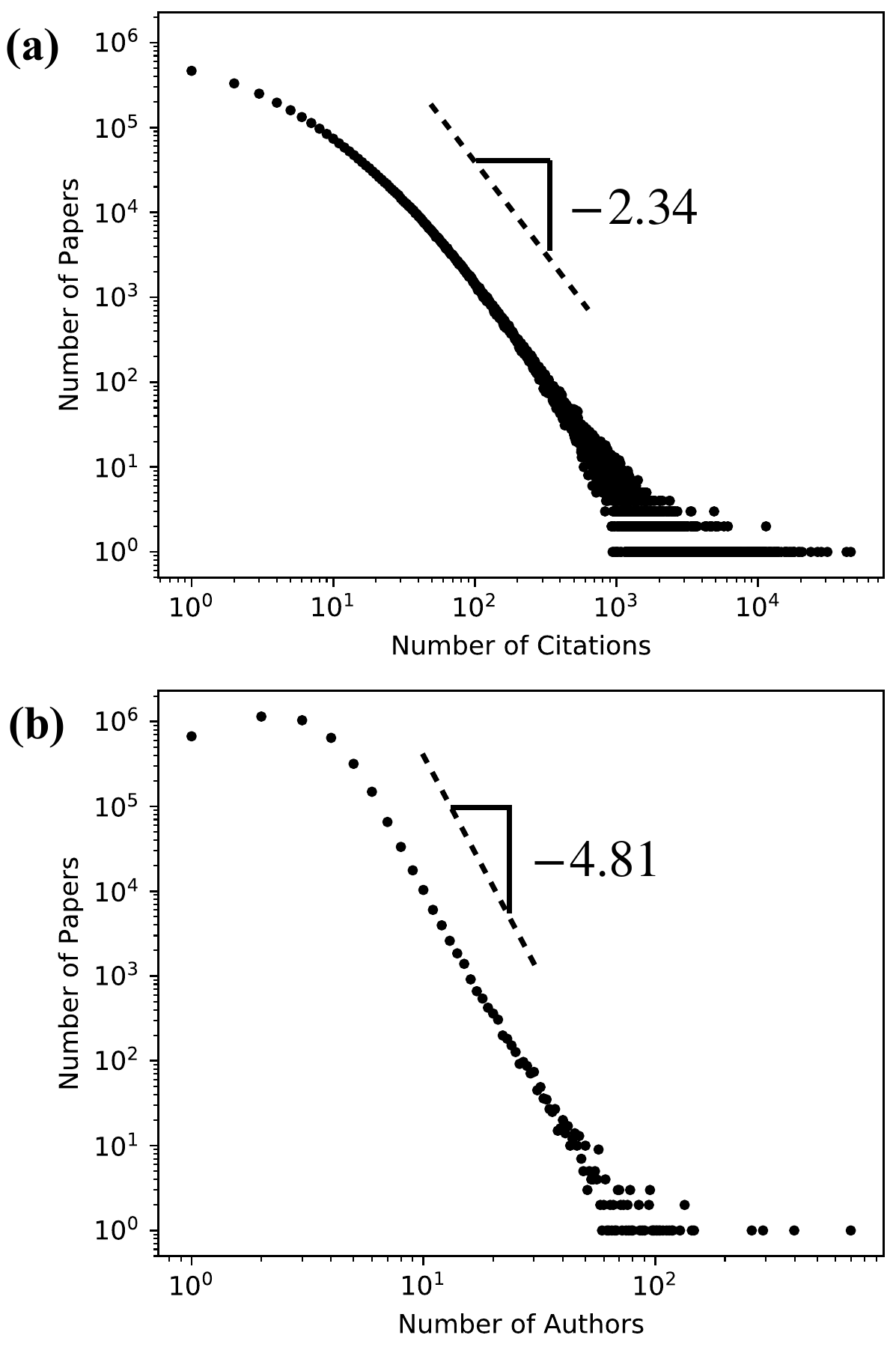}
\caption{Distribution of (a) number of citations and (b) number of authors for all papers from the DBLP bibliography from between 1940 and 2019. Dashed lines and indicated slopes represent fitted power-law exponents fitted within regions (a) 50 to 500 citations, and (b) 5 to 30 authors.}
\label{Fig1}
\end{figure}

Before examining the time-evolution of citation trends, I first discuss the general citation statistics of the dataset. FIG. \ref{Fig1}(a) shows the citation distribution of the entire dataset from 1940 to 2019. Publications that were never cited are not shown in the figure due to the logarithmic scaling of axes. The distribution of citations is qualitatively similar to that from the Physical Review citation network, as shown in \cite{redner2004citation, redner1998popular}. The distribution if citations is characterized by an overall curvature on a log-log plot with an apparent power-law regime between 50 and 500 citations. A power-law fit to the citation counts in this regime shows an exponent $-2.34$. The magnitude of this exponent is slightly less than the $-2.55$ of publications from the Physical Review journals reported in Ref. \cite{redner2004citation} and $-2.93$ of publications from high h-index chemists reported in Ref. \cite{peterson2010nonuniversal}. The smaller power-law exponent signifies a slower decay in the number of publications with many citations. This abundance of highly cited papers is likely a result of the larger number of publications produced in computer science compared to other scientific disciplines.

FIG. \ref{Fig1}(b) shows the distribution of the number of authors from all publications. The most common number of authors is two, followed by three, and then one author. Publications with two or three authors comprise 53.1\% of all publications in the dataset. The popularity of two and three author publications arises from the typical academic research environment where an advisor and student co-author research papers \cite{lee2000publication}. The distribution of publications with 5 or more authors seem to show a more consistent, power-law decay. A power-law fit to the range of authors from 5 to 30 results in an exponent $-4.81$, signifying a rapid decay in the number of publications with more than five authors. The proportion of publications with more than 30 authors in computer science is less than 0.05\%.

\begin{figure}[h]
\centering
\includegraphics[width=8.5cm]{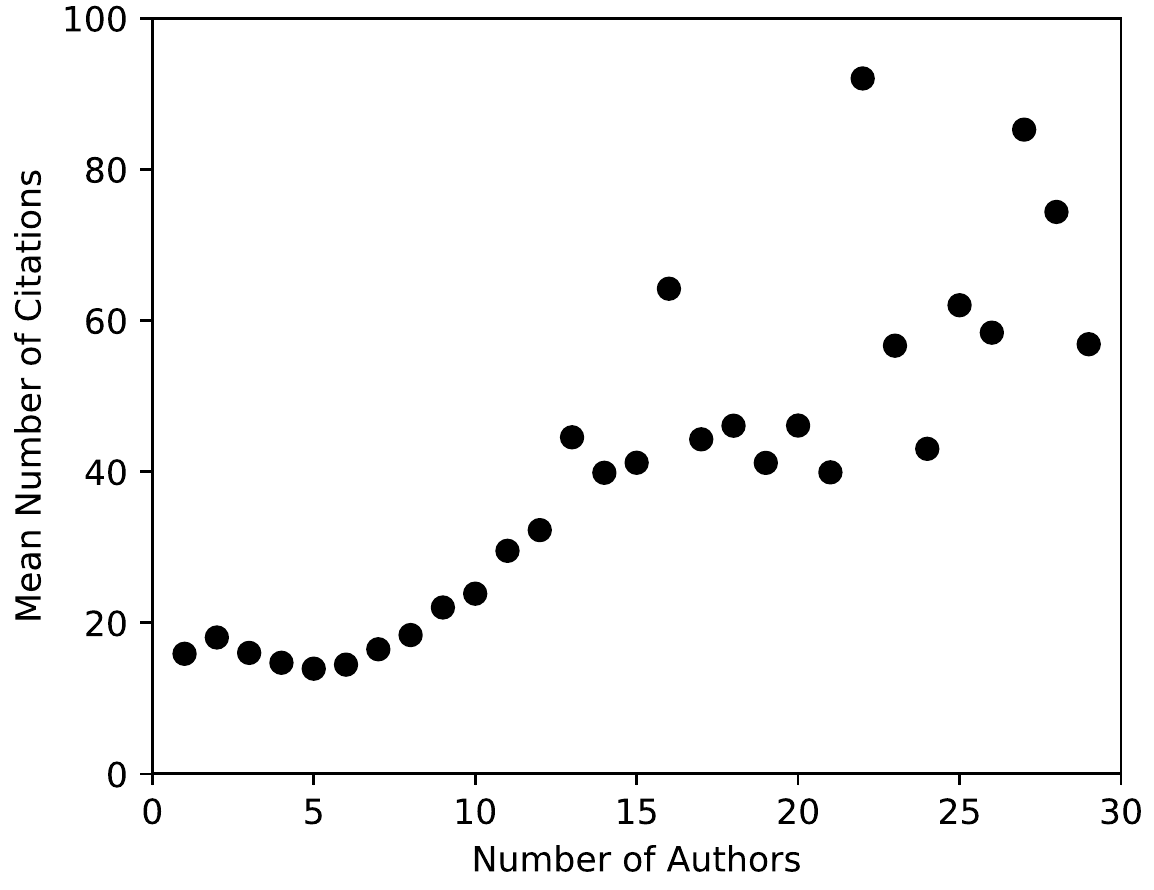}
\caption{The mean number of citations vs. number of authors for publications with less than 30 authors from the DBLP bibliography.}
\label{Fig2}
\end{figure}

An analysis of the relationship between the number of authors in a publication and the mean number of citations received shows that the two variables are strongly coupled. FIG. \ref{Fig2} shows the relation between the mean number of citations and authors from the entire dataset. Due to the scarcity of publications with greater than 30 publications, only publications with less than 30 authors are shown. There exists a strong correlation between the average number of citations and the number of authors. On average, publications with two authors are cited more compared to those with one or three to seven authors. Publications with more then eight authors on average have received more citations than those with fewer authors. The underlying reason behind the non-monotonic trend between citation and authorship is unclear, although the larger number of citations for works authored by more than seven researchers may be a result of the expanded scope and impact of \textit{larger projects} that lead to more citations.

Interestingly, the non-monotonic relation of FIG. \ref{Fig2} for publications with less than seven authors is qualitatively different from those observed in other fields. In applied ecology, the mean number of citations increases monotonically with the number of authors \cite{barlow2018extinction}. A bibliographic study in economics shows another trend where sole authorship appears to be the the important factor \cite{kuld2018rise}. Single authored publications receive less citations than co-authored publications, while there was no significant difference between publications with two to four authors.

\begin{figure}[h]
\centering
\includegraphics[width=8.5cm]{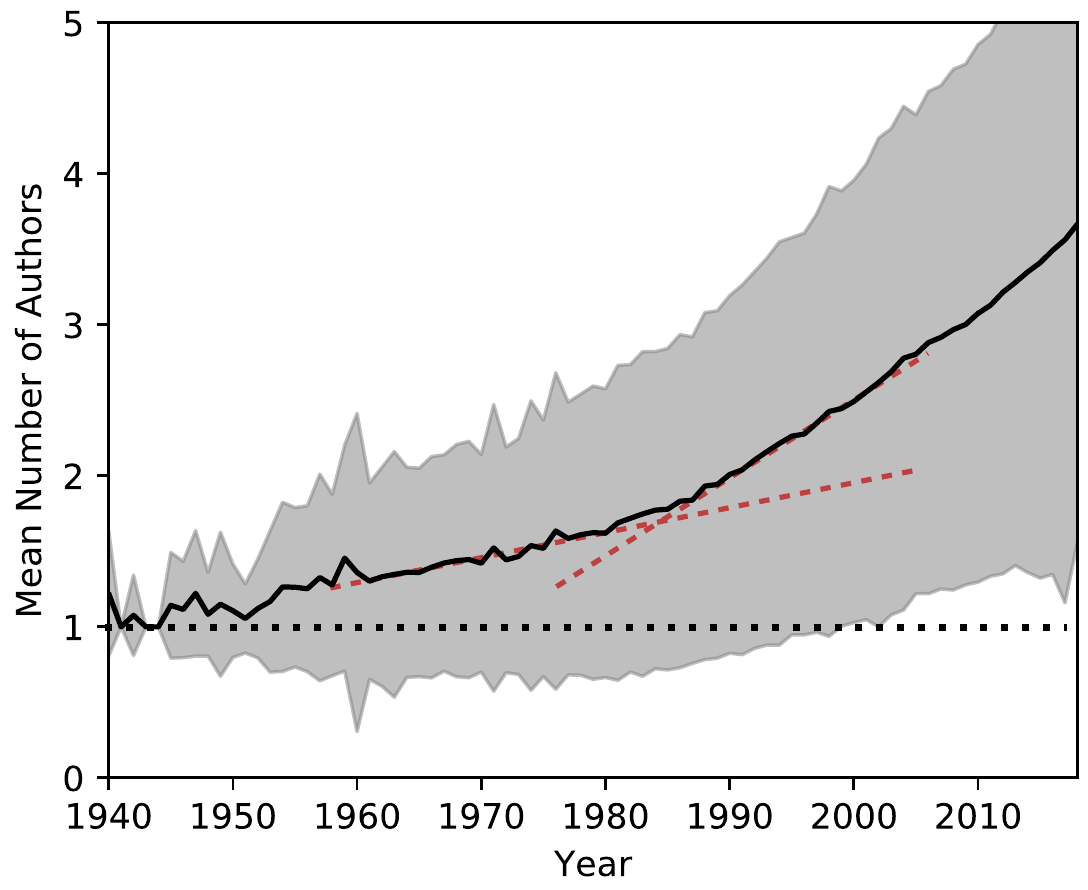}
\caption{Mean number of authors per publication vs. publication year. Grey shaded regions indicate a standard deviation above and below average. Black dotted line is drawn to guide the eye for one author. Red dashed lines are drawn to guide the eye to the change in slope from regions 1970 - 1985 to 1986 - 2000.}
\label{Fig3}
\end{figure}

Now I discuss how authorship has changed over the past six decades. Over the history of computer science, the number of authors per publication has increased quite consistently. FIG. \ref{Fig3} shows the average number of authors for all computer science publications as a function of the year of publication. Regions shaded in gray represent one standard deviation above and below the mean number of authors among all publications from that publication year. The growth in co-authorship has led to an increase from an average of one author in the 1940s to over three authors in 2010s. This constant increase in co-authorship is consistent with trends observed in other scientific disciplines \cite{behrens2010mathematics, kuld2018rise, fernandes2017evolution}; computer science is part of the universal rise in co-authorship and collaboration.

The trend in FIG. \ref{Fig3} shows an acceleration in the growth of co-authorship starting from the late 1980s, indicated by an increase in the slope. This marked increase in collaboration and the number of authors has been reported in previous scientometric studies. Rosenblat and Mobius argued that the advances in communication technologies and means of transportation brought researchers closer together and facilitated collaborative research \cite{rosenblat2004getting}. The advent of fax reduced the cost of communication in the 1980s and the wide availability of internet allowed rapid exchange of large documents and datasets in the 1990s. With the availability of collaborative research, however, came a large spread of mean authorship, as indicated by the widening of the gray region in FIG. \ref{Fig3} that represents a standard deviation above and below the mean. The spread in authorship indicates that publications with few authors have persisted for several decades, even after collaborative research gained popularity. Hence in present day there exists a diverse range of authorship in academic publications. Yet, single author publications have now become outside of a standard deviation away from the mean since the early 2000s illustrating that sole authorship has become rare in the 21st century.

\subsection{The Role of Single Author Papers in Computer Science Literature}
The trends in co-authorhship discussed in the previous section with FIG. \ref{Fig3} suggests a decline in single authored papers. In this subsection, I present a detailed study on the population and reception of single author papers over the past decades.

\begin{figure}[h]
\centering
\includegraphics[width=8.5cm]{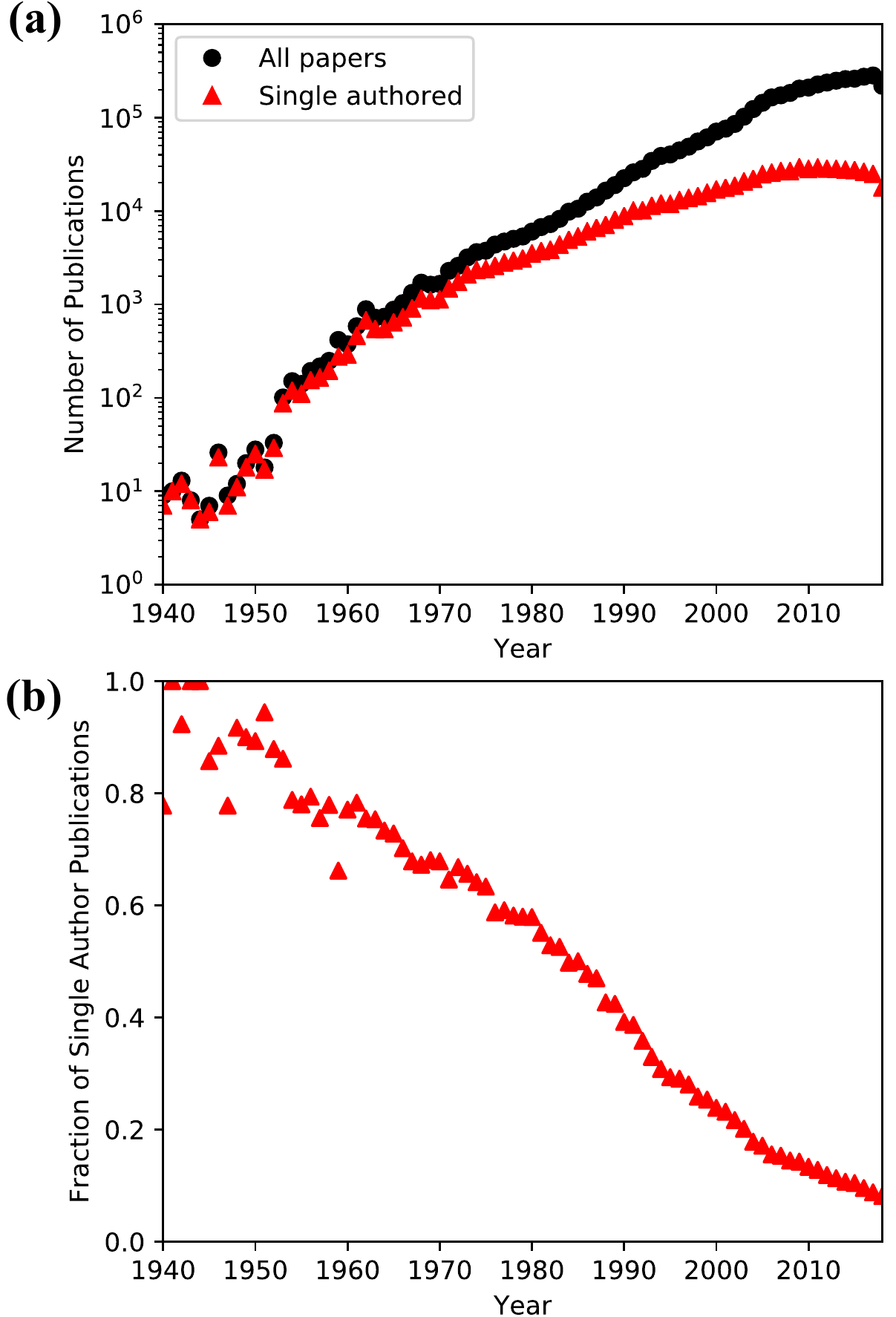}
\caption{(a) Number of all (black circles) and single author (red triangles) publications published each year. (b) Fraction of single author publications published each year.}
\label{Fig4}
\end{figure}

The total number of all publications and single authored publications in the DBLP database have both increased over time. FIG. \ref{Fig4} compares the population of these two publication groups over publication years from 1940 to 2019. The computer science literature has grown explosively from producing around 100 publications annually from the 1950s to publishing over 200,000 papers annually. With the growth of computer science, the number of single-authored works published every year rose as well. However, the two groups eventually branched from the 1970s and the upward trajectory of single authored publications reached a plateau in the mid 2000s (FIG. \ref{Fig4}(a)). 
 
FIG. \ref{Fig4}(a) showed that the publication count of single author publication is increasing slower than all publications. FIG \ref{Fig4}(b) shows the decline in population fraction of single author publications from each publication year. It is evident that the slow decline in single authored publication started in the 1940s and still persists today. Similar decaying trends have been reported in many other scientific disciplines \cite{abt2007future, behrens2010mathematics, barlow2018extinction}. Interestingly, other disciplines showed much faster exponential declines in publications with sole authors, eventually reporting near extinctions of single author publications by the 1990s. In computer science, however, the proportion of single authored publications in computer science has been declining slowly and roughly linearly. At 1990, single author publications still constituted 40\% of all computer science publications, and this value only fell below 10\% in the late 2010s. In 2018, 7.4\% of papers published were written by a sole author. In 1963 Price had predicted that single author papers will become extinct by the 1980s \cite{price1963solla}. This statement did not prove to be true as authors have continued to publish papers alone in the computer science field three decades past the 1980s. As of today, however, single author publications may become virtually defunct by the mid 2020s at current rate.

The overall demise of single author publication can be explained in several ways. In addition to the overall growth of interdisciplinary science that was enabled by improved transportation and communication technologies, a shift in scientific funding structure may be accountable for decrease in sole authorship. Studies in higher education have reported that the number of scientists in training, namely Masters students, Ph.D. students, and postdoctoral scholars, have increased over the past decades \cite{muller2012collaborating, salonius2012social}. Scientific studies published by scientists under training typically include the supervising lead researcher or principal investigator as authors. As a result, the relative number of studies conducted by financially and intellectually independent scientists has been declining.

\begin{figure}[h]
\centering
\includegraphics[width=8.5cm]{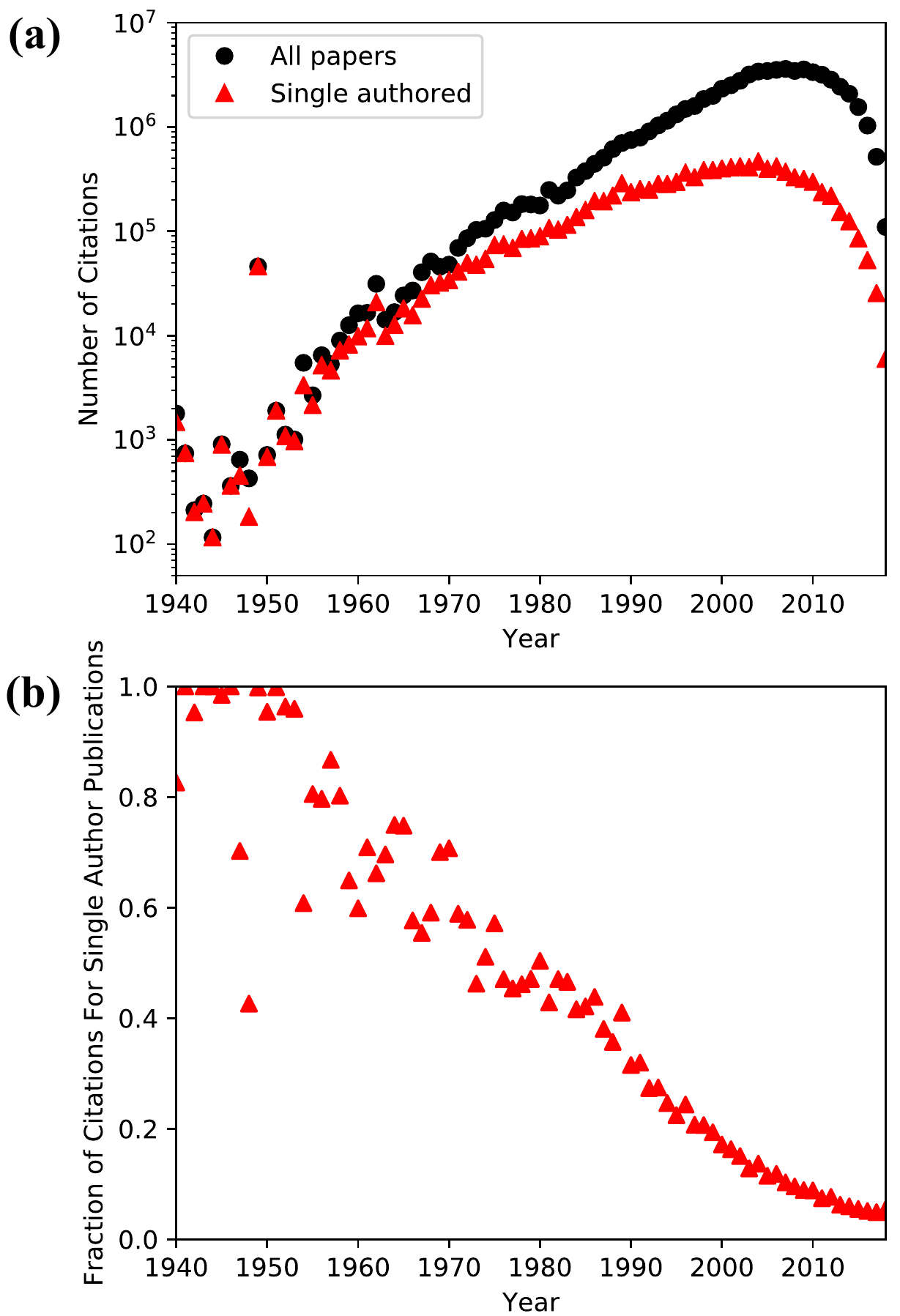}
\caption{(a) Number of citations received by all (black circles) and single author (red triangles) publications, and (b) fraction of citations received by single author publications as a function of publication year.}
\label{Fig5}
\end{figure}

A question closely related to the wane of single authored publications is whether 0single authored publications becoming less important in the computer science literature. That is, are single author publications not received as well compared to co-authored publications? To study the reception of single author publications, I show the number of citations received by single author publications in FIG. \ref{Fig5}(a). The figure shows the total number of citations for all and single authored publications for each paper publication year from 1940 to 2019. Over the past six decades, the total number of annual citations has increased rapidly for both all and single author publications. The increasing trend is strongly coupled to the sharp increase in the number of publications published each year as seen in FIG. \ref{Fig4}. The total number of citations reaches a peak at 2008, followed by a drop in the following years. This fall is due to the young age of publications.

Similar to the fall of population fraction seen in FIG. \ref{Fig4}(b), the fraction of citations received by single author publications have fallen over the past six decades. FIG. \ref{Fig5}(b) shows the fraction of citations given to single author papers published in each year. Though the trends seen in FIG. \ref{Fig5}(b) resemble that of population fractions, citation fractions of single author publications have declined faster than population fractions. Prior to 1970, single author publications received more than half of all citations. However, this fraction has rapidly decreased, especially in the late 1980s and 1990s. Among all citations of papers published in 2018, single author publications accounted for only 4.9\%, which is less than the population fraction of the same year (7.4\%). This result suggests that single author publications today are not received as well as other co-authored publications.

\begin{figure}[t]
\centering
\includegraphics[width=8.5cm]{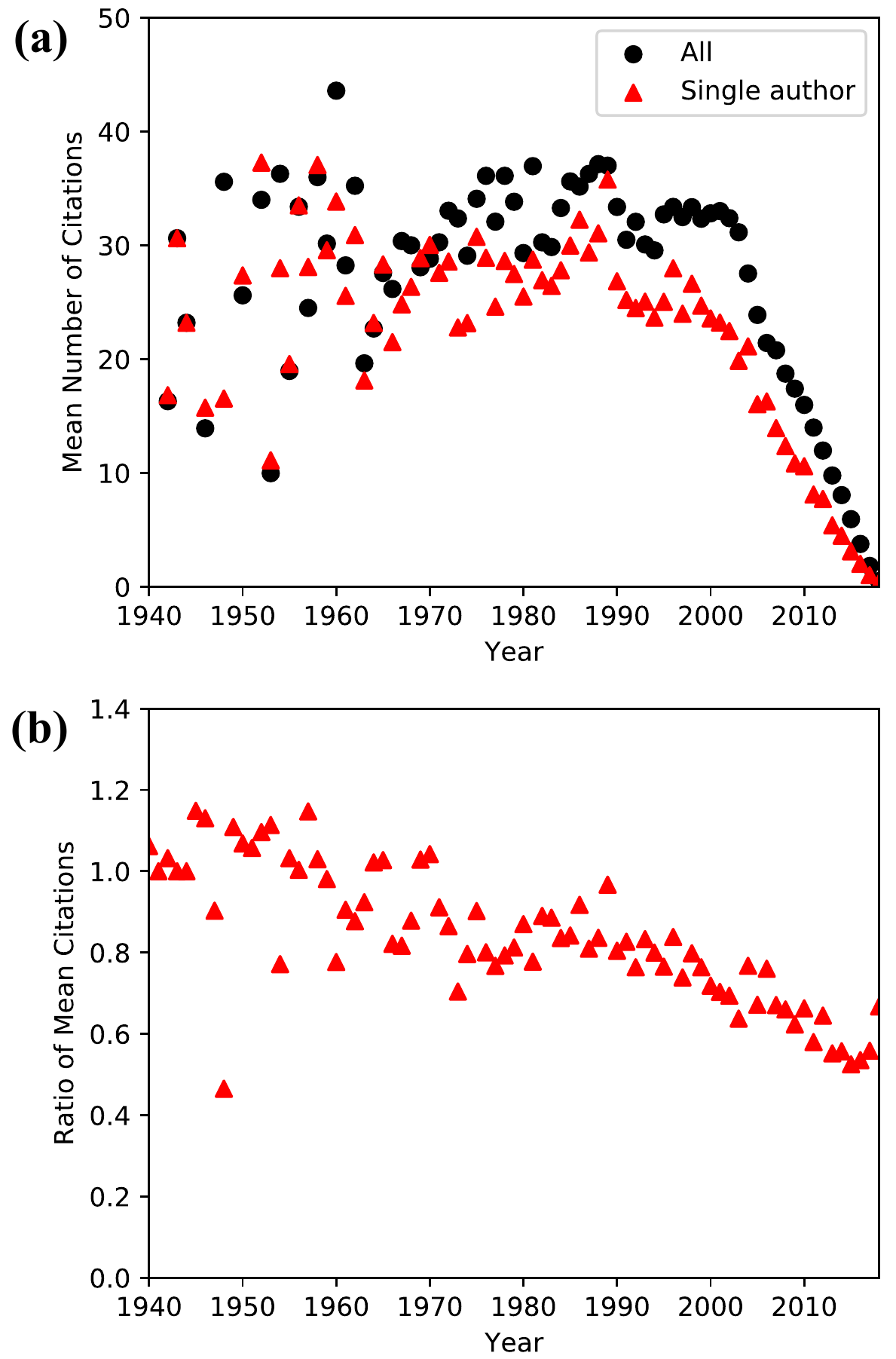}
\caption{(a) Mean number of citations received per publications from all (black circles) and single author (red triangles) publications, and (b) ratio of mean citations as a function of publication year.}
\label{Fig6}
\end{figure}

In FIG. \ref{Fig6}(a), I show the mean number of citations per publication to decouple the effects of decreasing population fraction and decreasing citation fractions from FIGs. \ref{Fig4} and \ref{Fig5}. The figure shows that the mean number of citations per publication is approximately constant between those published from 1970 to 2000. However, papers from the 21st century tend to have fewer publications on average. The relatively consistent number of citations for older publications shows that most publications eventually phase out from influence with age, which is consistent with the observation of Render \cite{redner2004citation}. Despite the occasional \textit{revival of old classics} or recognition of old \textit{discoveries}, publications in general gradually receive fewer citations as they age. The sharp decline in mean citation count for publications published after the early 2000s is again due to their young age. Furthermore, single author works published after 1980 constantly receive fewer citations per paper compared to the average of all works, conveying a decline in reception of single author works in the computer science literature.

A more detailed inspection of how well single author publications are cited on average relative to all publications is shown in FIG. \ref{Fig6}(b). The figure shows the ratio of mean number of citations of single author publications to the same mean of all publications published each year. Because this ratio decouples the effect of paper count, examining this ratio as a function of publication year allows a direct measurement of how well the single author publications are received by the computer science community. The overall downward trajectory seen in FIG. \ref{Fig6}(b) signifies that the average single author publication is cited less than the average co-authored publication. In other words, the role of single author publications in computer science literature has been diminishing throughout the past decades. This decay could be explained by the expanding scope of scientific research projects; interdisciplinary and large-scale research is difficult to be conducted by a sole author. This predicament may be due to the immense amount of domain knowledge required to pursue interdisciplinary research, or the computational resources required to conduct larger-scale projects. As a result, single author publications may find it challenging to meet the present day's standards for well-received research.

Although the number of citations is the most common and readily accessible metric for quantifying the importance and acceptance of a publication, citation count as a metric involves several shortcomings. First, all citations are weighed the same, regardless of whether the referencing paper is a widely cited publication or a paper that has few citations. Second, older publications with key discoveries eventually cease to receive more citations when textbooks or review papers summarizing such discoveries emerge, but citation count is unable to provide credit to works that are referred to in a review paper or a textbook. An ideal metric should be able to capture this cascade of information. Third, raw citation counts do not account for severe imbalances between scientific fields. The average number of publications and citations can vary significantly across scientific disciplines, and key publications from each field will inherently have disparate citation counts.

An alternative metric to circumvent such shortcomings of citation counts suggested by previous works in bibliometrics literature is Google's PageRank algorithm \cite{maslov2008promise, fiala2008pagerank}. PageRank is a graph analysis algorithm that outputs PageRank scores for each node, which signifies a probability distribution arising from a random walk \cite{page1999pagerank}. When applied to a citation network, the algorithm randomly surveys the network, following directed citation relations and updates scores to the visited publications (nodes). The property PageRank scores are able to ameliorate the aforementioned shortcomings of citation counts.  Naturally, well-cited publications have a higher probability of being visited, hence leading to a higher PageRank score. Furthermore, PageRank gives more weight to citations from highly cited publications. As a result, citing a paper increases the PageRank score of not only the referenced publication, but also papers that were cited by that work. Hence, a diminishing yet cascading provides credit to the works cited in textbooks or review papers. Finally, because the score is normalized to a probability distribution, comparison of PageRank scores allow better comparison across works from scientific disciplines compared to raw citation counts.

Here I have computed PageRank scores using $d = 1/2$ as the damping factor to examine the influence of single author publications in the DBLP dataset. The damping factor sets the probability of discontinuing a random walk and beginning a new search. The original PageRank algorithm used $d \approx 1/6$ based on the assumption that a typical user surfing the internet clicks six links in each website before starting from a new page \cite{page1999pagerank}. However, the average length of sequential citation links followed by the typical audience of scientific publications should considerably less than 6. Therefore, I have chosen a value $d=1/2$ based on the study by Maslov and Redner \cite{maslov2008promise}, which focused on the usage of PageRank in the context of citation networks. A higher damping factor means the that on average, a researcher follows a shorter series of citations.

\begin{figure}[b]
\centering
\includegraphics[width=8.5cm]{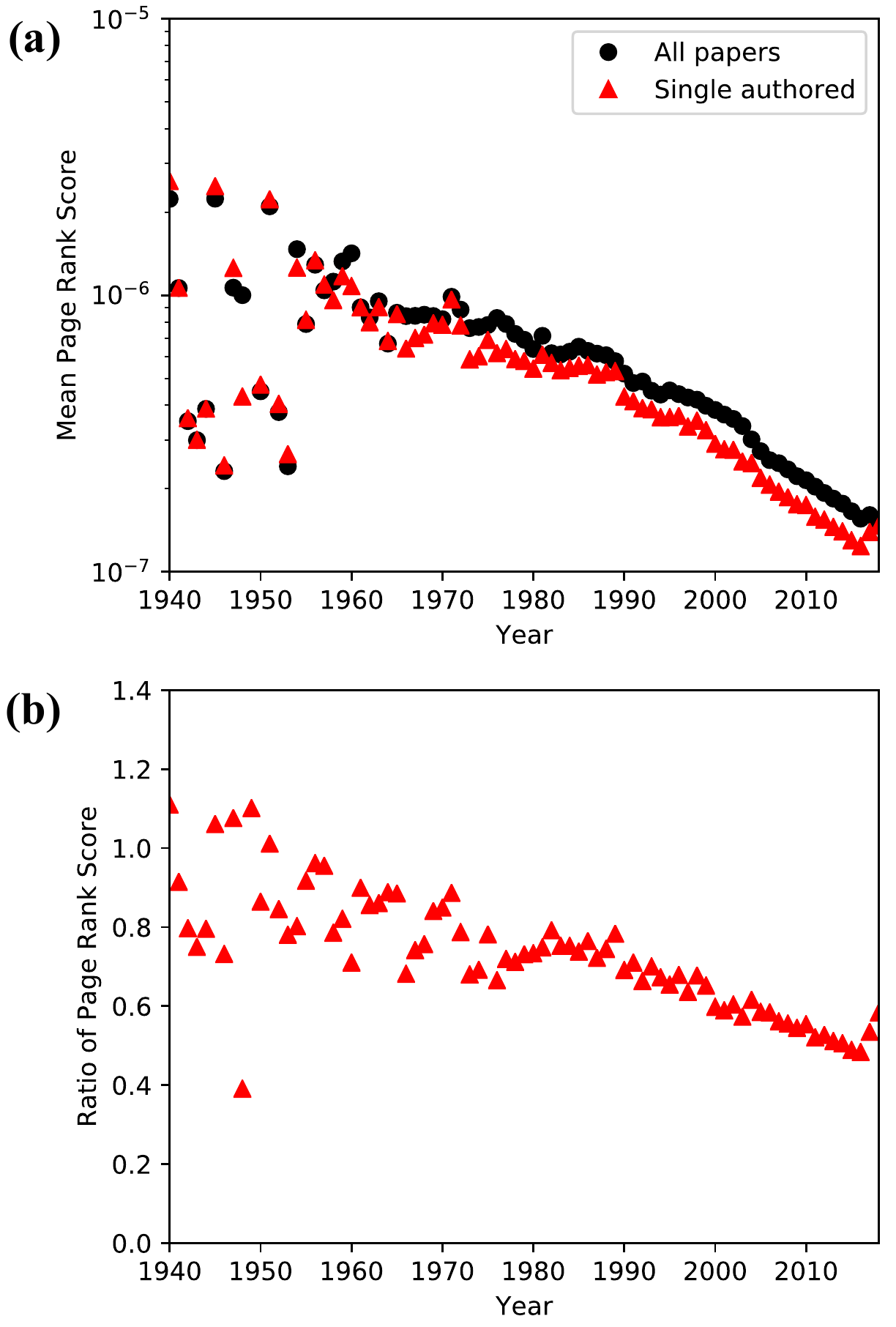}
\caption{(a) Mean number of PageRank scores of all (black circles) and single author (red triangles) publications by publication year, and (b) ratio of PageRank scores as a function of publication year.}
\label{Fig7}
\end{figure}

FIG. \ref{Fig7}(a) shows the average PageRank scores of the dataset for all and single author works published each year. Note that the page rank score decreases with publication year. This downwards trend is due to the directed, acyclic nature of the citation network; because citations point towards already published papers, a random walk in a citation network inevitably progresses towards older papers. Therefore, PageRank scores for older publication are higher on average, making analysis and comparison of PageRank scores only useful for comparing publications of the same age.

Focusing on publications published each year reveals the authorship dependence of PageRank scores. FIG. \ref{Fig7}(b) shows the ratio of mean PageRank scores between single author and all publications for each publication year. Since the 1960s, the average PageRank scores of single author publications have consistently been less than the mean of all publications. The downward trajectory further indicates that the influence of single author publications measured by PageRank scores is consistently declining. As of 2018, the average citation score of single author publications is only 60\% of that of the total average. This value is notably similar to the ratio of mean citations from FIG. \ref{Fig7}(b). In other words, the analysis of PageRank scores validates the diminishing role of single author publications seen from citation counts. Single author publications are dwindling not only in population fraction, but also in importance. 

\subsection{Potential Factors that Affect Citation Counts of Single Authored Publications}
The results in FIGs. \ref{Fig4}-\ref{Fig7} support that single author publications are decreasing both in the proportion as well as citation count. The poor citation performance of single author publications naturally brings up the question of whether single author publications lack quality. Does there exist a correlation between the number of authors and the quality of a publication? In addition to the citation relations, the DBLP bibliographic dataset mined via Arnetminer contains information regarding page count and the number of references in each publication \cite{tang2008arnetminer}. Here I assume that the length and reference count of a publication can serve as a first order approximation of the work's volume and significance and compare these two factors between single authored publications to the average of all publications.

\begin{figure}[h]
\centering
\includegraphics[width=8.5cm]{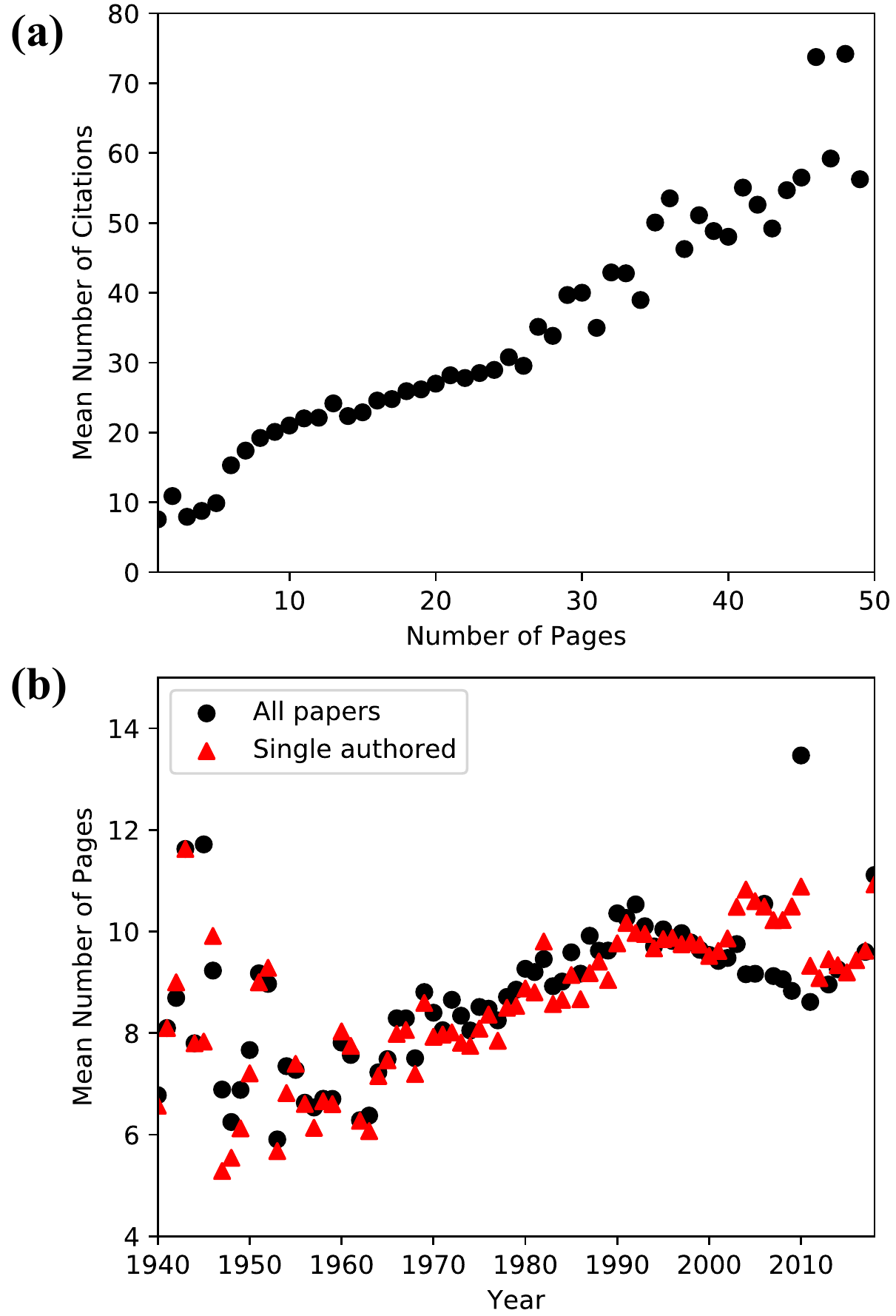}
\caption{(a) Mean number of citations received as a function of the page length of each publication. (b) Comparison of mean number of pages for all (black circles) and single author (red triangles) publications by publication year.}
\label{Fig8}
\end{figure}

First, I examine whether the citation count and the length of publication, measured in number of pages, are correlated. FIG \ref{Fig8}(a) illustrates this relationship for publication lengths ranging from 1 to 50 pages. Results for papers that are longer than 50 pages are not shown due to a scarcity of longer publications. On average, there exists a positive correlation between the number of citations and the length of the publication. This correlation can be due to several reasons. A longer publication may simply cover a larger scope or present in-depth derivations or results that may lead to well-supported and significant findings that are well cited. Alternatively, review papers tend to be longer and receive more citations than regular publications.  

Do sole authors produce papers that are shorter than collaborative projects? FIG. \ref{Fig8}(b) compares the number of pages for all publications and single author publications by publication year. In general, the length of computer science publications has increased from the around 7 pages in the 1950s to about 9 pages in the 1990s. However, it is noteworthy that the average number of pages of publications with sole authors is not significantly less than those from multiple authors over all years shown. In fact, single author publications produced in the 21st century tend to be longer than the average publication. Therefore, authorship and length of publication are in general, uncorrelated. A similar lack of correlation between authorship and publication length has been reported in the economics literature as well \cite{kuld2018rise}. In summary, these observations demonstrate that although publication length is correlated to the number of citations, single author publications are not shorter than co-authored publications in general and publication length is not the reason why single authored publications are cited less.

\begin{figure}[t]
\centering
\includegraphics[width=8.5cm]{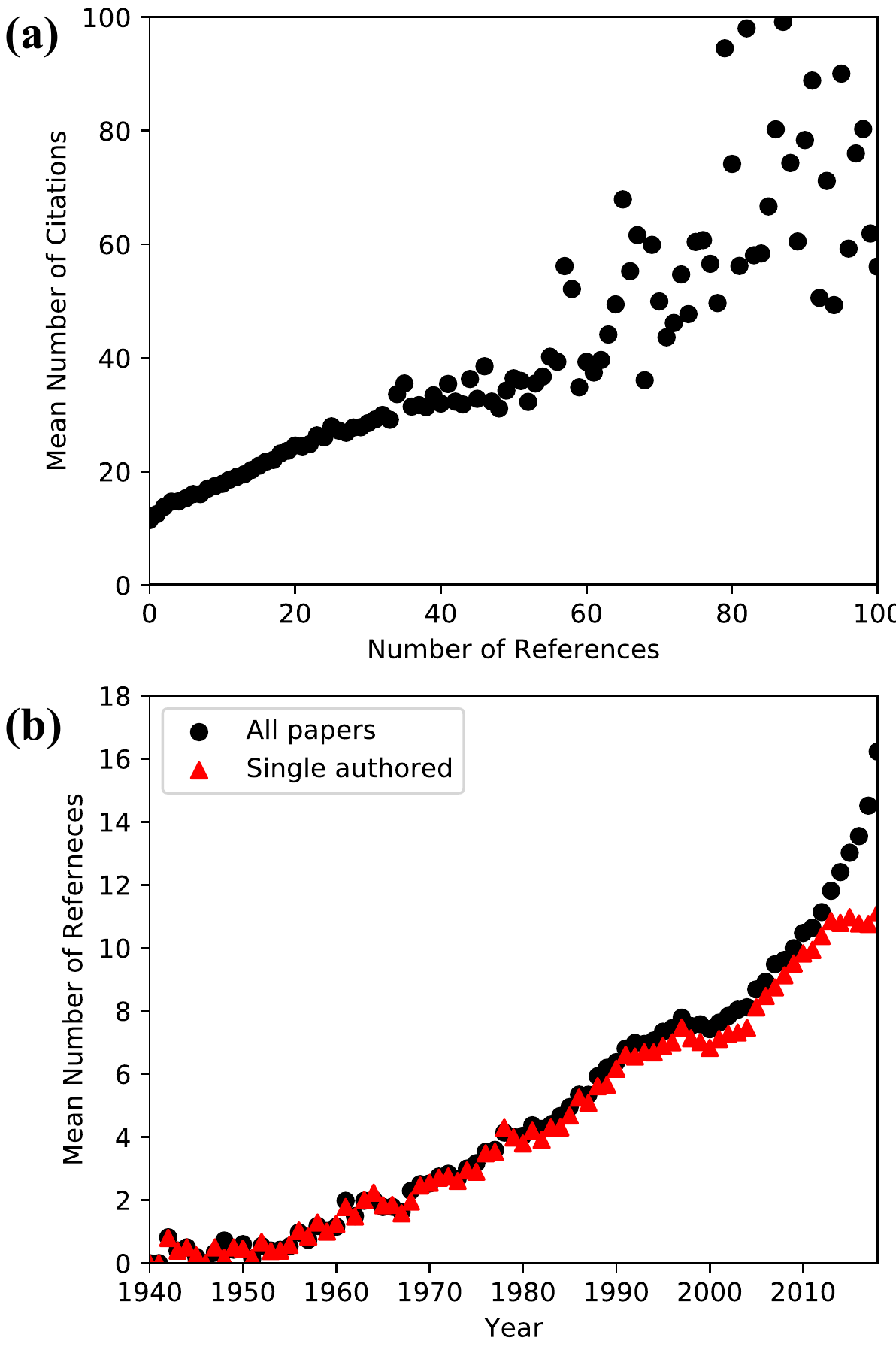}
\caption{(a) Mean number of citations received as a function of the number of references in each publication. (b) Comparison of mean number references for all (black circles) and single author (red triangles) publications by publication year.}
\label{Fig9}
\end{figure}

Another metric related to the volume of a publication is the number of references in a paper. Publications with more references tend to provide a broader and deeper introduction and thorough comparison between presented results and published works in literature. To prove the significance of the number of references, FIG. \ref{Fig9}(a) shows the overall between the number of references and the number of citations from all publications. Publications with over 100 references are not shown due to the scarcity of datapoints. Moreover, I note that the number of references includes only those within the citation network dataset. Therefore, references pointing to publications that are not in the DBLP bibliography, such as those in non-computer science related fields, are not included. Similar to the trend between length and citations in FIG. \ref{Fig8}(a), the number of references and citations have a positive correlation up to 50 references. Again, this result suggests that publications with a greater number of references are likely to be cited more within the discipline. 

Similar to how single author publications were not shorter than the average publication, there does not seem to exist a significant difference between the number of references included in single author publications and all publications. FIG. \ref{Fig9}(b) presents the the number of references in each publication group by publication year. Note that publications preceding 1950 may appear to have no references because only references pointing to other works present in the DBLP citation network are considered. A striking trend is the rapid growth in the mean number of references. The average number of references in both single authored and all publications have grown similarly up to 2010, with virtually no quantitative difference. However, the average number of references from single author publications have reached a plateau at 2010 while the average of all publications are continuing to grow. As of 2018, the average for all publications was approximately 16, whereas the average for single author publications was less than 11. The apparent branching post-2010 does not appear to fully explain why single author publications are cited less in the computer science literature. The underlying reason for this distinct branching is unclear; a detailed future study of comparison of other factors such as publication type such as journal article vs. conference proceeding and the propensity of single author publications to be submitted to either type may elucidate the cause of such branching. A qualitative difference between the types of publications produced by sole authors and publications that are co-authored may result in the current computer science literature.

\begin{figure}[h]
\centering
\includegraphics[width=8.5cm]{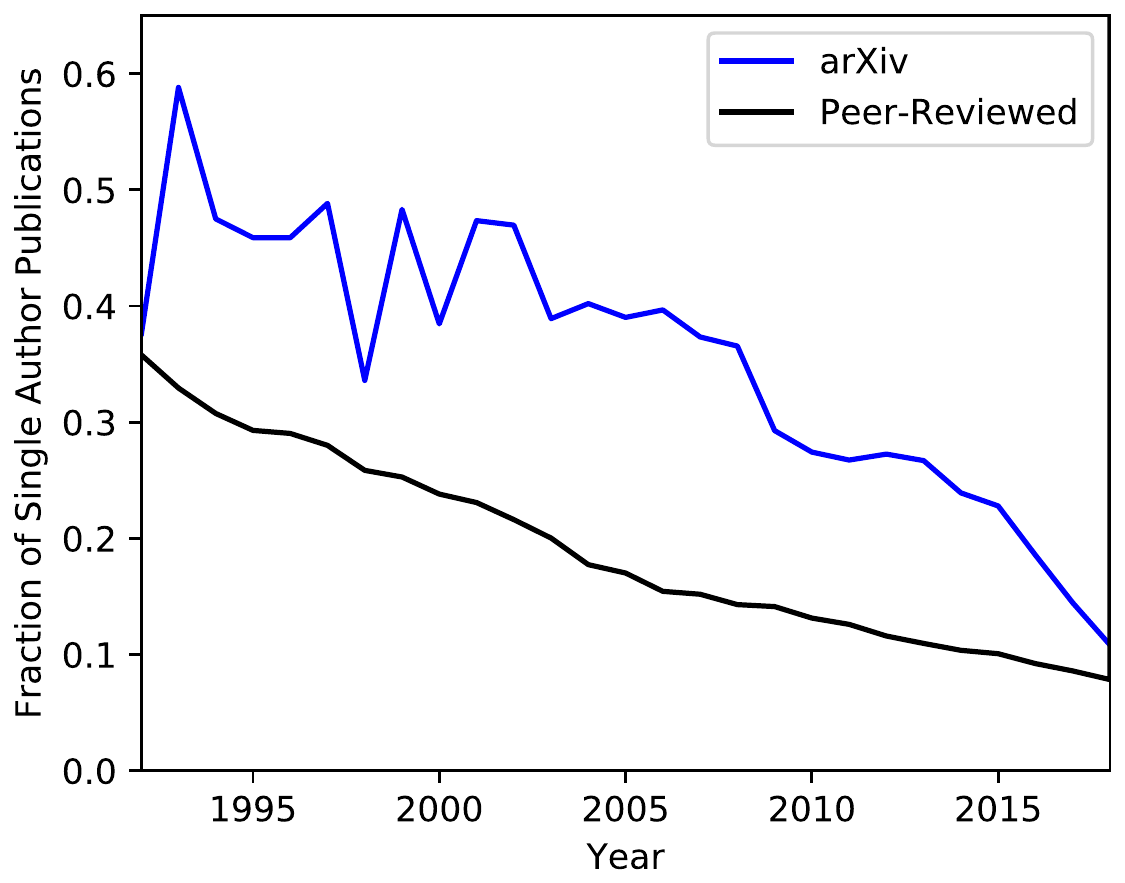}
\caption{Fraction of single author publications in arXiv (blue) and peer-reviewed publication venues (black) published each year.}
\label{Fig10}
\end{figure}

Finally, I examine whether single author publications have a proclivity to publish on the arXiv preprint server without peer review to obtain an understanding of publication submission trends. From a total of 4.1 million publications in the current dataset, 1.94\% of works are arXiv preprints tagged as computer science works. FIG. \ref{Fig10} shows time-evolution of the fraction of single author studies in arXiv and other peer-reviewed conference proceedings or journals since the inception of arXiv in 1992. The fraction of single author publications uploaded to arXiv have gradually decayed over time. Yet, the figure clearly shows that the fraction of single author publications uploaded to arXiv is significantly higher than the population fraction of single author publications published in peer-reviewed publication venues.

Why is arXiv more frequented by sole authors? Independent researchers may choose to not undergo the often rigorous peer-review process for several reasons. While publication fees and the time-consuming procedures of peer-reviews may deter single author publications from being published, single author publications uploaded to arXiv may simply lack scientific rigor or quality. Previous studies of articles uploaded to arXiv have reported plagiarized results \cite{sorokina2006plagiarism, bouville2008plagiarism, bechhoefer2007plagiarism} or contain fake data  \cite{labbe2013duplicate}. Bouville \cite{bouville2008plagiarism} states that the absence of a thorough plagiarism detection system in arXiv creates a huge driving force for plagiarism. Such a driving force may be stronger for publications with sole authors, where no other scientist can check for validity of data or possible text overlap. A future study examining authorship of dubious publications on arXiv would be able to provide insight on whether the number of authors is correlated to violation of science ethics.




\section{Conclusions}
This work examined the decline of single author publications in the computer science literature via the DBLP bibliographic dataset. In general, the rapid growth of the number of studies published in computer science has led to a growth in the number of single author publications. Along with the number of publications, however, the average number of authors in each publication has been continuously rising. This average increase in authorship naturally led to the decline in fractions of publications that are single authored. While single author publications comprised over 80\% of all papers published prior to 1950, only 7.4\% papers published in 2018 were authored by one researcher. The decrease in this ratio is due to a combination of several factors including the shift of scientific trends towards collaborative and interdisciplinary research, and the change in composition of scientific researchers, where there is an increased fraction of students who co-author publications with their advisors.

In conjunction with the decrease in fraction single author publications, the average number of citations received by single author publications has been declining over the past six decades. During the 1940s and 1950s, there was no palpable difference in the mean number of citations between single author publications and all publications. As of 2018, however, single author publications on average receive only 60\% of citations compared to the average of all works. An analysis of the holistic importance of single author publications in the citation network via PageRank scores reveals results consistent with mean citation count; on average, single author publications are losing its significance in the computer science literature.

Finally, the average length and reference count of single author publications were compared to those of all publications in the DBLP database. These metrics were examined to obtain a first-order approximation of the scope and volume of papers. While length and reference count both have positive correlation to the average number of times cited, there does not exist a large difference between the average and single author papers. Differences were only observed in the mean number of referernces for works published past 2010, which cannot fully explain the fewer citations single author publications have been receiving. Yet, noticeable differences in the popularity of arXiv as a means of of propagating science has been observed, where single author publications consitute a larger fraction of publications on arXiv compared to other peer-reviewed publication venues. Future studies that examine publication venues in detail may be able to extract further qualitatives difference between single author and co-authored publications. 

\nocite{*}

\bibliography{references.bib}

\end{document}